\documentclass[10pt,twocolumn,twoside] {IEEEtran}
\usepackage{amsmath}
\usepackage{times}
\usepackage{amsbsy,color}
\usepackage{latexsym}
\usepackage{amssymb}
\usepackage{graphicx}
\usepackage{subfigure}
\usepackage{amsfonts}
\usepackage{epsfig}
\allowdisplaybreaks %
\begin{document}

\title{ Distributed Compressed Estimation for Wireless Sensor Networks Based on Compressive Sensing }
\author{Songcen~Xu*,~
        Rodrigo C. de~Lamare,~\IEEEmembership{Senior Member,~IEEE,}
        and~H. Vincent~Poor,~\IEEEmembership{Fellow,~IEEE}  \vspace{-1em}
\thanks{S. Xu* is with the Department of Electronics, University of York, YO10 5DD York, U.K.
(e-mail: songcen.xu@york.ac.uk).}
\thanks{R. C. de Lamare is with CETUC / PUC-Rio, Brazil and Department of Electronics, University of York, U.K. (e-mail: rodrigo.delamare@york.ac.uk).}
\thanks{H. V. Poor is with the Department of Electrical Engineering, Princeton University,
Princeton NJ 08544 USA (e-mail: poor@princeton.edu).}
\thanks{EDICS: NET-DISP, NET-ADEG, NET-GRPH, NET-SPRS}}


\maketitle

\begin{abstract}
This letter proposes a novel distributed compressed estimation
scheme for sparse signals and systems based on compressive sensing
techniques. The proposed scheme consists of compression and
decompression modules inspired by compressive sensing to perform
distributed compressed estimation. A design procedure is also
presented and an algorithm is developed to optimize measurement
matrices, which can further improve the performance of the proposed
distributed compressed estimation scheme. Simulations for a wireless
sensor network illustrate the advantages of the proposed scheme and
algorithm in terms of convergence rate and mean square error
performance.
\end{abstract}


\IEEEpeerreviewmaketitle

\section{Introduction}

\IEEEPARstart{D}{istributed} signal processing algorithms are of
great importance for statistical inference in wireless networks and
applications such as wireless sensor networks (WSNs)
\cite{Lopes2,Xu,Xu1,Cattivelli3}. Distributed processing techniques
deal with the extraction of information from data collected at nodes
that are distributed over a geographic area \cite{Lopes2}. In this
context, for each node a set of neighbor nodes collect and process
their local information, and transmit their estimates to a specific
node. Then, each specific node combines the collected information
together with its local estimate to generate improved estimates.

In many scenarios, the unknown parameter vector to be estimated can
be sparse and contain only a few nonzero coefficients. Many
algorithms have been developed in the literature for sparse signal
estimation
\cite{Chen,Lorenzo,Angelosante,Chouvardas,Mateos,Lorenzo1,Zhaoting,Sayin,intspl,jiospl,Rcdl1,jidfstap,Rcdl2,barc}.
However, these techniques are designed to take into account the full
dimension of the observed data, which increases the computational
cost, slows down the convergence rate and degrades mean square error
(MSE) performance.

Compressive sensing (CS) \cite{Donoho,Candes} has recently received
considerable attention and been successfully applied to diverse
fields, e.g., image processing \cite{Romberg}, wireless
communications \cite{Bajwa} and MIMO radar \cite{Yao}. The theory of
CS states that an $S$--sparse signal $\boldsymbol \omega_{0}$ of
length $M$ can be recovered exactly with high probability from
$\mathcal{O}(S\log M)$ measurements. Mathematically, the vector
$\bar{\boldsymbol \omega}_{0}$ with dimension $D \times 1$ that
carries sufficient information about $\boldsymbol \omega_{0}$ ($D\ll
M$) can be obtained via a linear model \cite{Candes} \vspace{-0.5em}
\begin{equation}
{\bar{\boldsymbol {\omega}}}_0=\boldsymbol\Phi\boldsymbol{\omega}_0
\vspace{-0.5em}
\end{equation}
where $\boldsymbol\Phi\in R^{D\times M}$ is the measurement matrix.

The application of CS to WSNs has been recently investigated in
\cite{Bajwa,Wei}, \cite{Baron,Quer}. A compressive wireless sensing
scheme was developed in \cite{Bajwa} to save energy and bandwidth,
where CS is only employed in the transmit layer. In \cite{Wei}, a
greedy algorithm called precognition matching pursuit was developed
for CS and used at sensors and the fusion center to achieve fast
reconstruction. However, the sensors are assumed to capture the
target signal perfectly with only measurement noise. The work of
\cite{Baron} introduced a theory for distributed CS based on jointly
sparse signal recovery. However, in \cite{Baron} CS techniques are
only applied to the transmit layer, whereas distributed CS in the
estimation layer has not been widely investigated. A sparse model
that allows the use of CS for the online recovery of large data sets
in WSNs was proposed in \cite{Quer}, but it assumes that the sensor
measurements could be gathered directly, without an estimation
procedure. In summary, prior work has focused on signal
reconstruction algorithms in a distributed manner but has not
considered both compressed transmit strategies and estimation
techniques.

In this work, we focus on the design of an approach that exploits
lower dimensions, reduces the required bandwidth, and improves the
convergence rate and the MSE performance. Inspired by CS, we
introduce a scheme that incorporates compression and decompression
modules into the distributed estimation procedure. In the
compression module, we compress the unknown parameter $\boldsymbol
\omega_{0}$ into a lower dimension. As a result, the estimation
procedure is performed in a compressed dimension. After the
estimation procedure is completed, the decompression module recovers
the compressed estimator into its original dimension using an
orthogonal matching pursuit (OMP) algorithm \cite{Tropp,Pati,Davis}.
We also present a design procedure and develop an algorithm to
optimize the measurement matrices, which can further improve the
performance of the proposed scheme. Specifically, we derive an
adaptive stochastic gradient recursion to update the measurement
matrix. Simulation results illustrate the performance of the
proposed scheme and algorithm against existing techniques.

This paper is organized as follows. Section II describes the system
model. In Section III, the proposed distributed compressed
estimation scheme is introduced. The proposed measurement matrix
optimization is illustrated in Section IV. Simulation results are
provided in Section V. Finally, we conclude the paper in Section VI.

{\bf Notation}: We use boldface uppercase letters to denote matrices
and boldface lowercase letters to denote vectors. We use
$(\cdot)^{-1}$ to denote the inverse operator, $(\cdot)^H$ for
conjugate transposition and $(\cdot)^*$ for complex conjugate.
\vspace{-0.8em}

\section{System Model and Problem Statement}
\vspace{-0.4em} A wireless sensor network (WSN) with N nodes, which
have limited processing capabilities, is considered with a partially
connected topology. A diffusion protocol is employed although other
strategies, such as incremental \cite{Lopes1} and consensus
\cite{Xie} could also be used. A partially connected network means
that nodes can exchange information only with their neighbors as
determined by the connectivity topology. In contrast, a fully
connected network means that, data broadcast by a node can be
captured by all other nodes in the network \cite{Bertrand}. At every
time instant $i$, the sensor at each node $k$ takes a scalar
measurement $d_k(i)$ according to \vspace{-0.4em}
\begin{equation}
{d_k(i)} = {\boldsymbol {\omega}}_0^H{\boldsymbol x_k(i)}
+{n_k(i)},~~~ \label{desired signal12} i=1,2, \ldots, \textrm{I} ,
\end{equation}
where ${\boldsymbol x_k(i)}$ is the $M \times 1$ input signal vector
with zero mean and variance $\sigma_{x,k}^2$, ${n_k(i)}$ is the
noise at each node with zero mean and variance $\sigma_{n,k}^2$.
From (\ref{desired signal12}), we can see that the measurements for
all nodes are related to an unknown parameter vector ${\boldsymbol
{\omega}}_0$ with size $M \times 1$ that should be estimated by the
network. We assume that ${\boldsymbol {\omega}}_0$ is a sparse
vector with $S \ll M$ non-zero coefficients. The aim of such a
network is to compute an estimate of ${\boldsymbol{\omega}}_0$ in a
distributed fashion, which minimizes the cost function
\begin{equation}
{J_{\omega}({\boldsymbol \omega})} = \sum_{k=1}^{N}{\mathbb{E}\{ |{
d_k(i)}- {\boldsymbol \omega}^H{\boldsymbol x_k(i)}|^2}\} ,
\end{equation}
where $\mathbb{E}\{\cdot\}$ denotes expectation. Distributed
estimation of ${\boldsymbol{\omega}}_0$ is appealing because it
provides robustness against noisy measurements and improved
performance as reported in \cite{Lopes2,Lopes1,Xie}. To solve this
problem, a cost-effective technique is the adapt--then--combine
(ATC) diffusion strategy \cite{Lopes2}
\begin{equation}
\left\{\begin{array}{ll}
{\boldsymbol \psi}_k(i)= {\boldsymbol \omega}_k(i)+{\mu}_k {\boldsymbol x_k(i)}\big[{d_k(i)}-{\boldsymbol \omega}_k^H(i){\boldsymbol x_k(i)}\big]^*,\\
\ \\
{\boldsymbol {\omega}}_k(i+1)= \sum\limits_{l\in \mathcal{N}_k}
c_{kl} \boldsymbol\psi_l(i),
\end{array}
\right.
\end{equation}
where $\mathcal{N}_k$ indicates the set of neighbors for node $k$,
${\boldsymbol \psi}_k(i)$ is the local estimator of node $k$,
$|\mathcal{N}_k|$ denotes the cardinality of $\mathcal{N}_k$ and
$c_{kl}$ is the combination coefficient, which is calculated with
respect to the Metropolis rule
\begin{equation}
\left\{\begin{array}{ll} c_{kl}= \frac{1}
{\max(|\mathcal{N}_k|,|\mathcal{N}_l|)},\ \ \ \ \ \ \ \ \  \ \ \ \ \ \ \ \ \     $if\  $k\neq l$\  \ are\  linked$\\
c_{kl}=0,              \ \ \ \ \ \ \ \ \ \  \ \ \ \ \ \ \ \ \ \ \ \ \ \ \ \ \ \ \ \ \ \ \ $for\  $k$\  and\  $l$\ not\  linked$\\
c_{kk} = 1 - \sum \limits_{l\in \mathcal{N}_k / k} c_{kl}, \ \ \ \ \
\ \ \ \ \ \ \ \ \ \ \ \ $for\  $k$\ =\ $l$$
\end{array}
\right.
\end{equation}
and should satisfy
\begin{equation}
\sum\limits_{l} c_{kl} =1 , l\in \mathcal{N}_k \forall k .
\end{equation}
Existing distributed sparsity-aware estimation strategies, e.g.,
\cite{Chen,Lorenzo,Angelosante}, are designed using the full
dimension signal space, which reduces the convergence rate and
degrades the MSE performance. In order to improve performance,
reduce the required bandwidth and optimize the distributed
processing, we incorporate at each node of the WSN the proposed
distributed compressed estimation scheme based on CS techniques,
together with a measurement matrix optimization algorithm.
\vspace{-0.5em}

\section{Proposed Distributed Compressed Estimation Scheme}

\begin{figure}[!htb]
\begin{center}
\vspace{-1em}
\def\epsfsize#1#2{1.0\columnwidth}
\epsfbox{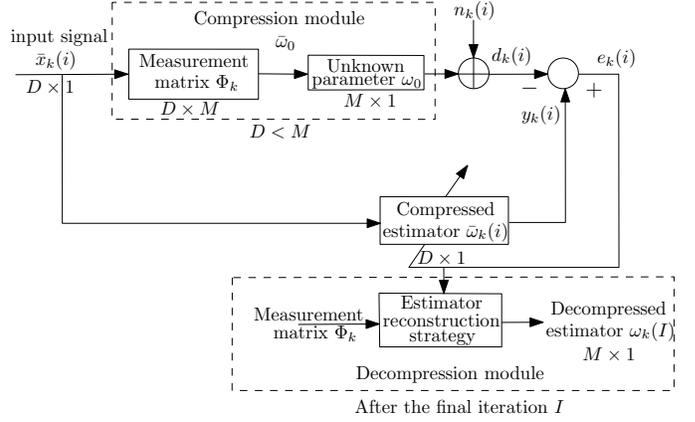} \vspace{-1.8em} \caption{\footnotesize
Proposed Compressive Sensing Modules} \vspace{-2.5em} \label{fig2}
\end{center}
\end{figure}

In this section, we detail the proposed distributed compressed
estimation (DCE) scheme based on CS. The proposed scheme, depicted
in Fig. \ref{fig2}, employs compression and decompression modules
inspired by CS techniques to perform distributed compressed
estimation. In the proposed scheme, at each node, the sensor first
observes the $M \times 1$ vector ${\boldsymbol x}_k(i)$, then with
the help of the $D\times M$ measurement matrix obtains the
compressed version ${\bar{\boldsymbol x}}_k(i)$, and performs the
estimation of $\boldsymbol \omega_{0}$ in the compressed domain. In
other words, the proposed scheme estimates the $D\times 1$ vector
$\bar{\boldsymbol\omega}_0$ instead of the $M\times 1$ vector
$\boldsymbol \omega_0$, where $D\ll M$ and the $D$--dimensional
quantities are designated with an overbar. At each node, a
decompression module employs a $D\times M$ measurement matrix
$\boldsymbol\Phi_k$ and a reconstruction algorithm to compute an
estimate of $\boldsymbol \omega_{0}$. One advantage for the DCE
scheme is that fewer parameters need to be transmitted between
neighbour nodes.

\begin{figure}[!htb]
\begin{center}
\vspace{-1em}
\def\epsfsize#1#2{1.0\columnwidth}
\epsfbox{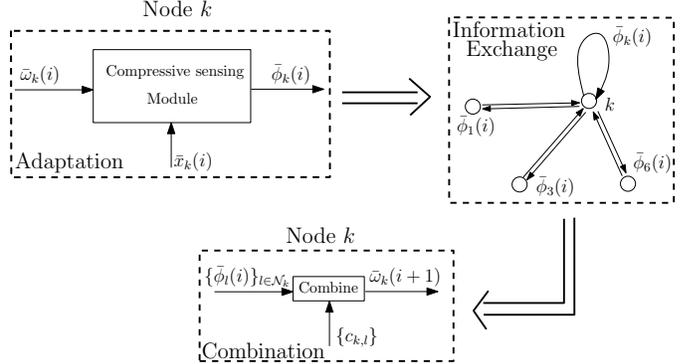} \vspace{-1.5em} \caption{\footnotesize
Proposed DCE Scheme} \vspace{-1.5em} \label{fig3}
\end{center}
\end{figure}

We start the description of the proposed DCE scheme with the scalar
measurement $d_k(i)$ given by
\begin{equation}
{d_k(i)} = {\bar{\boldsymbol {\omega}}}_0^H{\bar{\boldsymbol
x}_k(i)} +{n_k(i)},~~~ \label{desired signal} i=1,2, \ldots,
\textrm{I} ,
\end{equation}
where ${\bar{\boldsymbol {\omega}}}_0=\boldsymbol\Phi_k\boldsymbol
{\omega}_0$ and $\bar{\boldsymbol x}_k(i)$ is the $D\times 1$ input
signal vector. This operation is depicted in Fig. \ref{fig2} as the
compression module.

Fig. \ref{fig3} illustrates the proposed DCE scheme. The scheme can
be divided into three steps:
\begin{itemize}
\item Adaptation
\end{itemize}
In the adaptation step, at each time instant $i$=1,2, . . . , I,
each node $k$=1,2, \ldots, N, generates a local compressed estimator
${\boldsymbol {\bar{\psi}}}_k(i)$ through
\begin{equation}
{\boldsymbol {\bar{\psi}}}_k(i)=
\bar{\boldsymbol\omega}_k(i)+\mu_k(i)e_k^*(i)\bar{\boldsymbol
x}_k(i),
\end{equation}
where $e_k(i)=d_k(i)-\bar{\boldsymbol\omega}_k^H(i){\bar{\boldsymbol
x}_k(i)}$ and $\mu_k(i)=\frac{\mu_0}{\bar{\boldsymbol
x}_k^H(i)\bar{\boldsymbol x}_k(i)}$.
\begin{itemize}
\item Information exchange
\end{itemize}
Given the network topology structure, only the local compressed
estimator ${\boldsymbol {\bar{\psi}}}_k(i)$ will be transmitted
between node $k$ and all its neighbor nodes. The measurement matrix
$\boldsymbol\Phi_k$ will be kept locally.
\begin{itemize}
\item Combination
\end{itemize}
At each time instant $i$=1,2, . . . , I, the combination step starts
after the information exchange is finished. Each node will combine
the local compressed estimators from its neighbor nodes and itself
through
\begin{equation}
\bar{{\boldsymbol {\omega}}}_k(i+1)= \sum\limits_{l\in
\mathcal{N}_k} c_{kl}\boldsymbol{\bar{\psi}}_l(i),
\end{equation}
to compute the updated compressed estimator $\bar{\boldsymbol
{\omega}}_k(i+1)$.

After the final iteration $I$, each node will employ the OMP
reconstruction strategy to generate the decompressed estimator
$\boldsymbol\omega_k(I)$. Other reconstruction algorithms can also
be used. The decompression module described in Fig. \ref{fig2}
illustrates the details. In summary, during the DCE procedure, only
the local compressed estimator ${\boldsymbol {\bar{\psi}}}_k(i)$
will be transmitted over the network resulting in a reduction of the
number of parameters to be transmitted from $M$ to $D$. The proposed
DCE scheme is given in Table \ref{table1}.

The computational complexity of the proposed DCE scheme is
$O(NDI+ND^3)$, where $N$ is the number of nodes in the WSN and $I$
is the number of time instants. The distributed NLMS algorithm has a
complexity $O(NMI)$, while the complexity of the sparse diffusion
NLMS algorithm \cite{Lorenzo} is $O(3NMI)$. For the distributed
compressive sensing algorithm of \cite{Wei}, the computational
complexity is $O(NMI+ND^3I)$. In the proposed DCE scheme, only the
local compressed estimator ${\boldsymbol {\bar{\psi}}}_k(i)$ with
$D$ parameters will be transmitted through the network, which means
the transmission requirement is greatly reduced as compared with the
standard schemes that transmit ${\boldsymbol {\psi}}_k(i)$ with $M$
parameters.

\begin{table}[!htb]\scriptsize
\vspace{-1.5em} \centering \caption{{The Proposed DCE Scheme}}
\begin{tabular}{l}\hline
Initialize: ${\boldsymbol {\bar{\omega}}}_k(1)$=0\\
For each time instant $i$=1,2, . . . , I-1\\
\ \ \ \ For each node $k$=1,2, \ldots, N\\
\ \ \ \ \ \ \ \ \ \ ${\boldsymbol {\bar{\psi}}}_k(i)= \bar{\boldsymbol\omega}_k(i)+\mu(i)e_k^*(i)\bar{\boldsymbol x}_k(i)$\\
\ \ \ \ \ \ \ \ \ \ where $e_k(i)=d_k(i)-\bar{\boldsymbol\omega}_k^H(i){\bar{\boldsymbol x}_k(i)}$,\\
\ \ \ \ \ \ \ \ \ \ ${d_k(i)} = {\bar{\boldsymbol {\omega}}}_0^H{\bar{\boldsymbol x}_k(i)} +{n_k(i)}=(\boldsymbol\Phi_k\boldsymbol {\omega}_0)^H{\bar{\boldsymbol x}_k(i)} +{n_k(i)}$\\
\ \ \ \ \ \ \ \ \ \ and $\boldsymbol\Phi_k$ is the $D\times M$ random measurement matrix\\
\ \ \ \ end\\
\ \ \ \ For each node $k$=1,2, \ldots, N\\
\ \ \ \ \ \ \ \ \ \ $\bar{{\boldsymbol {\omega}}}_k(i+1)= \sum\limits_{l\in \mathcal{N}_k} c_{kl}\boldsymbol{\bar{\psi}}_l(i)$\\
\ \ \ \ end\\
end\\
After the final iteration $I$\\
For each node $k$=1,2, \ldots, N\\
\ \ \ \ ${\boldsymbol {\omega}}_k(I)= f_{\textrm{OMP}}\{{\boldsymbol {\bar{\omega}}}_k(I)\}$ \\
\ \ \ \ where ${\boldsymbol {\omega}}_k(I)$ is the final decompressed estimator.\\
end\\
\hline
\end{tabular}
\label{table1} \vspace{-2.3em}
\end{table}

\section{Measurement Matrix Optimization}

To further improve the performance of the proposed DCE scheme, an
optimization algorithm for the design of the measurement matrix
$\boldsymbol\Phi_k(i)$, which is now time--variant, is developed
here. Unlike prior work \cite{Yao,Yao1}, this optimization is
distributed and adaptive. Let us consider the cost function
\begin{equation}
\begin{split}
\mathcal{J}&= \mathbb{E}\{|e_k(i)|^2\}= \mathbb{E}\{|d_k(i)-y_k(i)|^2\} \\
&=\mathbb{E}\{|d_k(i)|^2\}-\mathbb{E}\{d_k^*(i)y_k(i)\}-\mathbb{E}\{d_k(i)y_k^*(i)\}\\
&\ \ \ +\mathbb{E}\{|y_k(i)|^2\},\label{CF}
\end{split}
\end{equation}
where $y_k(i)=\bar{\boldsymbol\omega}_k^H(i)\bar{\boldsymbol
x}_k(i)$. To minimize the cost function, we need to compute the
gradient of $\mathcal{J}$ with respect to $\boldsymbol\Phi_k^*(i)$
and equate it to a null vector, i.e.,
$\nabla\mathcal{J}_{\boldsymbol\Phi_k^*(i)}={\boldsymbol 0}$. As a
result, only the first three terms in (\ref{CF}) need to be
considered. Taking the first three terms of (\ref{CF}) we arrive at
\begin{align}
&\mathbb{E}\{|d_k(i)|^2\}-\mathbb{E}\{d_k^*(i)y_k(i)\}-\mathbb{E}\{d_k(i)y_k^*(i)\}\notag\\
&=\mathbb{E}\{|\boldsymbol\omega_0^H\boldsymbol\Phi_k^H(i)\bar{\boldsymbol x}_k(i)+n_k(i)|^2\}\notag\\
&\ \ \ -\mathbb{E}\{(\boldsymbol\omega_0^H\boldsymbol\Phi_k^H(i)\bar{\boldsymbol x}_k(i)+n_k(i))^*y_k(i)\} \notag\\
&\ \ \ -\mathbb{E}\{(\boldsymbol\omega_0^H\boldsymbol\Phi_k^H(i)\bar{\boldsymbol x}_k(i)+n_k(i))y_k^*(i)\} \notag\\
&=\mathbb{E}\{|\boldsymbol\omega_0^H\boldsymbol\Phi_k^H(i)\bar{\boldsymbol x}_k(i)|^2\}+\mathbb{E}\{(\boldsymbol\omega_0^H\boldsymbol\Phi_k^H(i)\bar{\boldsymbol x}_k(i))^*n_k(i)\}\notag\\
&\ \ \ +\mathbb{E}\{(\boldsymbol\omega_0^H\boldsymbol\Phi_k^H(i)\bar{\boldsymbol x}_k(i))n_k^*(i)\}+\mathbb{E}\{|n_k(i)|^2\}\notag\\
&\ \ \ -\mathbb{E}\{(\boldsymbol\omega_0^H\boldsymbol\Phi_k^H(i)\bar{\boldsymbol x}_k(i))^*y_k(i)\}-\mathbb{E}\{n_k^*(i)y_k(i)\}\notag\\
&\ \ \
-\mathbb{E}\{(\boldsymbol\omega_0^H\boldsymbol\Phi_k^H(i)\bar{\boldsymbol
x}_k(i))y_k^*(i)\}-\mathbb{E}\{n_k(i)y_k^*(i)\}.\label{fcf}
\end{align}
Because the random variable $n_k(i)$ is statistically independent
from the other parameters and has zero mean, (\ref{fcf}) can be
further simplified as
\begin{align}
&\mathbb{E}\{|d_k(i)|^2\}-\mathbb{E}\{d_k^*(i)y_k(i)\}-\mathbb{E}\{d_k(i)y_k^*(i)\}\notag\\
&=\mathbb{E}\{|\boldsymbol\omega_0^H\boldsymbol\Phi_k^H(i)\bar{\boldsymbol x}_k(i)|^2\}+\sigma_{n,k}^2-\mathbb{E}\{(\boldsymbol\omega_0^H\boldsymbol\Phi_k^H(i)\bar{\boldsymbol x}_k(i))^*y_k(i)\}\notag\\
&\ \ \
-\mathbb{E}\{(\boldsymbol\omega_0^H\boldsymbol\Phi_k^H(i)\boldsymbol
x_k(i))y_k^*(i)\}.
\end{align}
Then, we have
\begin{equation}
\nabla\mathcal{J}_{\boldsymbol\Phi_k^*(i)}= \boldsymbol
R_k(i)\boldsymbol\Phi_k(i)\boldsymbol
R_{\boldsymbol\omega_0}-\boldsymbol P_k(i),\label{cf1}
\end{equation}
where $\boldsymbol R_k(i)=\mathbb{E}\{\bar{\boldsymbol
x}_k(i)\bar{\boldsymbol x}_k^H(i)\}$, $\boldsymbol
R_{\boldsymbol\omega_0}=\mathbb{E}\{\boldsymbol \omega_0\boldsymbol
\omega_0^H\}$ and $\boldsymbol
P_k(i)=\mathbb{E}\{y_k^*(i)\bar{\boldsymbol x}_k(i)\boldsymbol
\omega_0^H\}$. Equating (\ref{cf1}) to a null vector, we obtain
\begin{equation}
\boldsymbol R_k(i)\boldsymbol\Phi_k(i)\boldsymbol
R_{\boldsymbol\omega_0}-\boldsymbol P_k(i)=\boldsymbol 0,
\end{equation}
\begin{equation}
\boldsymbol\Phi_k(i)= \boldsymbol R_k^{-1}(i)\boldsymbol
P_k(i)\boldsymbol R_{\boldsymbol\omega_0}^{-1}.\label{ex}
\end{equation}
The expression in (\ref{ex}) cannot be solved in closed--form
because $\boldsymbol\omega_0$ is an unknown parameter. As a result,
we employ the previous estimate $\bar{\boldsymbol\omega}_k(i)$ to
replace $\boldsymbol\omega_0$. However,
$\bar{\boldsymbol\omega}_k(i)$ and $\boldsymbol\Phi_k(i)$ depend on
each other, thus, it is necessary to iterate (\ref{ex}) with an
initial guess to obtain a solution. In particular, we replace the
expected values with instantaneous values. Starting from
(\ref{cf1}), we use instantaneous estimates to compute
\begin{equation}
\hat{\boldsymbol R}_k(i)=\bar{\boldsymbol x}_k(i)\bar{\boldsymbol
x}_k^H(i),
\end{equation}
\begin{equation}
\hat{\boldsymbol R}_{\boldsymbol\omega_0}=\boldsymbol
\omega_0\boldsymbol \omega_0^H
\end{equation}
and
\begin{equation}
\hat{\boldsymbol P}_k(i)=y_k^*(i)\bar{\boldsymbol x}_k(i)\boldsymbol
\omega_0^H.
\end{equation}
According to the method of steepest descent \cite{Haykin1}, the
updated parameters of the measurement matrix $\boldsymbol\Phi_k(i)$
at time $i+1$ are computed by using the simple recursive relation
\begin{align}
&\boldsymbol\Phi_k(i+1)=\boldsymbol\Phi_k(i)+\eta[-\nabla\mathcal{J}_{\boldsymbol\Phi_k^*(i)}]\notag\\
&=\boldsymbol\Phi_k(i)+\eta[ \hat{\boldsymbol P}_k(i)-\hat{\boldsymbol R}_k(i)\boldsymbol\Phi_k(i) \hat{\boldsymbol R}_{\boldsymbol\omega_0}]\\
&=\boldsymbol\Phi_k(i)+\eta[y_k^*(i)\bar{\boldsymbol
x}_k(i)\boldsymbol\omega_0^H-\bar{\boldsymbol
x}_k(i)\bar{\boldsymbol
x}_k^H(i)\boldsymbol\Phi_k(i)\boldsymbol\omega_0\boldsymbol\omega_0^H].\notag
\end{align}
where $\eta$ is the step size and $\boldsymbol\omega_0$ is the
$M\times1$ unknown parameter vector that must be estimated by the
network. Then, the parameter vector $\bar{\boldsymbol\omega}_k(i)$
is used to reconstruct the estimate of $\boldsymbol\omega_0$ as
follows
\begin{equation}
\boldsymbol\omega_{re_k}(i)=
f_{\textrm{OMP}}\{\bar{\boldsymbol\omega}_k(i)\},
\end{equation}
where the operator $f_{\textrm{OMP}}\{\cdot\}$ denotes the OMP
reconstruction algorithm. Note that other reconstruction algorithms
could also be employed. Replacing $\boldsymbol\omega_0$ by
$\boldsymbol\omega_{re_k}(i)$, we arrive at the expression for
updating the measurement matrix described by
\begin{align}
\boldsymbol\Phi_k(i+1)&=\boldsymbol\Phi_k(i)+\eta\big[y_k^*(i)\bar{\boldsymbol x}_k(i)\boldsymbol\omega_{re_k}^H(i)\notag\\
&-\bar{\boldsymbol x}_k(i)\bar{\boldsymbol
x}_k^H(i)\boldsymbol\Phi_k(i)\boldsymbol\omega_{re_k}(i)\boldsymbol\omega^H_{re_k}(i)\big].
\end{align}
The computational complexity of the proposed scheme with measurement
matrix optimization is $O(NDI+ND^3I)$. \vspace{-0.5em}
%
%
%

\section{Simulations}
We assess the proposed DCE scheme and the measurement matrix
optimization algorithm in a WSN application, where a partially
connected network with $N$ = 20 nodes is considered. We compare the
proposed DCE scheme with uncompressed schemes, including the
distributed NLMS (dNLMS) algorithm (normalized version of
\cite{Lopes2}), sparse diffusion NLMS algorithm \cite{Lorenzo},
sparsity-promoting adaptive algorithm \cite{Chouvardas}, and the
distributed compressive sensing algorithm \cite{Wei}, in terms of
MSE performance. Note that other metrics such as mean-square
deviation (MSD) could be used but result in the same performance
hierarchy between the analyzed algorithms.

The input signal is generated as ${\boldsymbol x}_k(i)=[x_k(i)\ \ \
x_k(i-1)\ \ \ ...\ \ \ x_k(i-M+1)]^T$  and
$x_k(i)=u_k(i)+\alpha_kx_k(i-1)$, where $\alpha_k$ is a correlation
coefficient and $u_k(i)$ is a white noise process with variance
$\sigma^2_{u,k}= 1-|\alpha_k|^2$, to ensure the variance of
${\boldsymbol x}_k(i)$ is $\sigma^2_{x,k}= 1$. The compressed input
signal is obtained by $\bar{\boldsymbol x}_k(i)=\boldsymbol\Phi_k
{\boldsymbol x}_k(i)$. The measurement matrix $\boldsymbol\Phi_k$ is
an i.i.d. Gaussian random matrix that is kept constant. The noise
samples are modeled as complex Gaussian noise with variance
$\sigma^2_{n,k}=
0.001$. 
The unknown $M\times 1$ parameter vector ${\boldsymbol \omega_0}$
has sparsity $S$, where $M$=50, $D$=10 and $S$=3. The step size
$\mu_0$ for the distributed NLMS, distributed compressive sensing,
sparse diffusion LMS and the proposed DCE algorithms is 0.45. The
parameter that controls the shrinkage in \cite{Lorenzo} is set to
0.001. For \cite{Chouvardas}, the number of hyperslabs equals 55 and
the width of the hyperslabs is 0.01.

\begin{figure}[!htb]
\begin{center}
\def\epsfsize#1#2{0.75\columnwidth}
\vspace{-1.2em}
\epsfbox{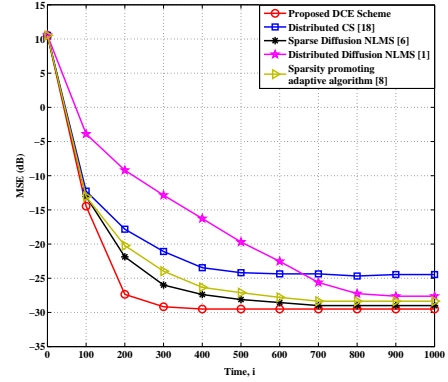} \vspace{-1.5em} \caption{\footnotesize
MSE performance against time} \vspace{-2.0em} \label{fig4}
\end{center}
\end{figure}

Fig. \ref{fig4} illustrates the comparison between the DCE scheme
with other existing algorithms, without the measurement matrix
optimization. It is clear that, when compared with the existing
algorithms, the DCE scheme has a significantly faster convergence
rate and a better MSE performance. These advantages consist in two
features: the compressed dimension brought by the proposed scheme
and CS being implemented in the estimation layer. As a result, the
number of parameters for transmission in the network is
significantly reduced.

\begin{figure}[!htb]
\begin{center}
\def\epsfsize#1#2{0.75\columnwidth}
\epsfbox{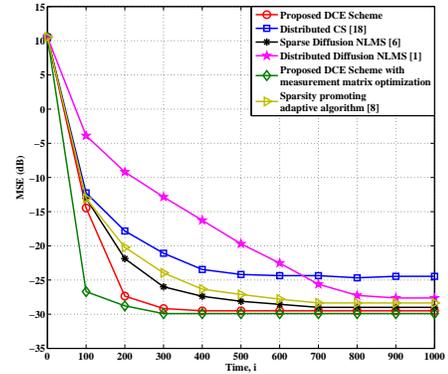} \vspace{-1.5em} \caption{\footnotesize
MSE performance against time with measurement matrix optimization}
\vspace{-1.5em} \label{fig5}
\end{center}
\end{figure}

In the second scenario, we employ the measurement matrix
optimization algorithm to in the DCE scheme. The parameter $\eta$
for the measurement matrix optimization algorithm is set to 0.08 and
all other parameters remain the same as in the previous scenario. In
Fig. \ref{fig5}, we observe that with the help of the measurement
matrix optimization algorithm, DCE can achieve a faster convergence
when compared with DCE without the measurement matrix optimization.

\begin{figure}[!htb]
\begin{center}
\def\epsfsize#1#2{0.75\columnwidth}
\vspace{-0.5em}
\epsfbox{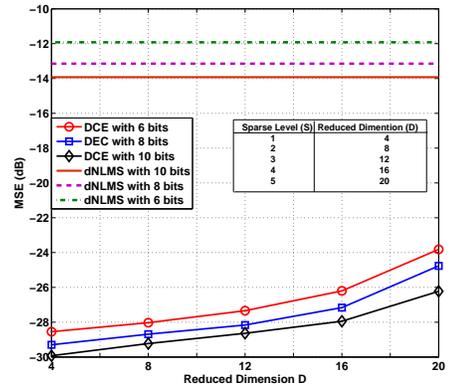} \vspace{-1.5em} \caption{\footnotesize
MSE performance against reduced dimension $D$ for different levels
of resolution in bits per coefficient} \vspace{-1.5em} \label{fig6}
\end{center}
\end{figure}

In the third scenario, we compare the DCE scheme with the
distributed NLMS algorithm with different levels of resolution in
bits per coefficient, reduced dimension $D$ and sparsity level $S$.
The x-axis stands for the reduced dimension $D$ and their
corresponding sparsity level $S$ can be found in Fig. \ref{fig6}. In
Fig. \ref{fig6}, it is clear that with the increase of the sparsity
level $S$ the MSE performance degrades. In addition, the MSE
performance will increase when the transmission has more bits per
coefficient. For the DCE scheme, the total number of bits required
for transmission is $D$ times the number of bits per coefficient,
whereas for the distributed NLMS algorithm it is $M$ times the
number of bits per coefficient. A certain level of redundancy is
required between the sparsity level and the reduced dimension due to
the error introduced by the estimation procedure.

\section{Conclusions}
\vspace{-0.05em} We have proposed a novel DCE scheme and algorithms
for sparse signals and systems based on CS techniques and a
measurement matrix optimization. In the DCE scheme, the estimation
procedure is performed in a compressed dimension. The results for a
WSN application show that the DCE scheme outperforms existing
strategies in terms of convergence rate, reduced bandwidth and MSE
performance.

{
\bibliographystyle{IEEEbib}
\bibliography{reference}}

\end{document}